# Adapting a 3D scanning water phantom for use in brachytherapy dosimetry

Rachael Wilks[1,2,3][0000-0002-2110-9676], Samuel Peet[1,2,4][0000-0003-3300-6495], Tanya Kairn[1,2,3,4][0000-0002-2136-6138], and Scott Crowe[1,2,3,4][0000-0001-7028-6452]

[1] Royal Brisbane & Women's Hospital, Herston QLD 4029, Australia
[2] Herston Biofabrication Institute, Herston QLD 4029, Australia
[3] University of Queensland, St. Lucia QLD 4072, Australia
[4] Queensland University of Technology, Brisbane QLD 4001, Australia
sb.crowe@gmail.com

**Abstract.** In external beam radiotherapy, 3D scanning water phantoms are the gold standard for obtaining relative dosimetry data. These phantoms, consisting of a water tank and mechanical arm, along with accompanying software, are designed to acquire dose profiles along and orthogonal to the beam axis. In brachytherapy, the acquisition of analogous dose profiles is more difficult, and is generally achieved with complex custom-built phantoms or chemical dosimeters such as film or gel. In this study, a low-cost 3D-printed jig was designed and fabricated within a clinical department, to allow precise brachytherapy dose measurements using a PTW BeamScan water phantom. Specifically, this jig allowed applicators to be suspended securely and reproducibly within the water phantom. Protocols were developed to relate the scanning system coordinates to the physical source position, and to obtain isodose planes both parallel and radial to the source axis. The developed solution has the potential to be used for physical verification of TG43 dose calculation parameters (e.g. anisotropy functions), the characterization of dose for a single dwell position in a complex applicator containing non-water equivalent materials, or the collection of point dose measurements for treatments incorporating multiple dwell positions or catheters.

## 1 Introduction

It is essential to validate the dose distribution of a radiotherapy treatment to ensure the intended outcome for the patient. This is particularly important for hypofractionated, high dose treatment techniques such as stereotactic ablative radiotherapy and brachytherapy, where an error in treatment calculation or delivery is less likely to be noticed over the short course of treatment, or captured by planning margins.

Although methods for performing in vivo measurements have been developed for brachytherapy procedures, the uncertainties inherent in these challenging measurements [1] have meant that clinical treatment dose calculation verification has tended to rely on independent dose calculations rather than physical measurements. Independent dose calculation approaches have been adequate until recently, due to the wide adoption of the AAPM's TG43 dose calculation algorithm, which does not account for different material densities surrounding the brachytherapy source [2]. However, now that institutions are developing customized applicators of non-water equivalent materials [3-5], there is a movement towards model-based calculation algorithms [6] resulting in the requirement for treatment dose verification methods capable of taking density heterogeneities into account.

The high dose gradients that make brachytherapy so attractive for highly conformal treatments also make relative dosimetry measurements problematic. Detector-source positioning uncertainty is one of the main contributing factors to this measurement difficulty, where a displacement of 2.5 mm could lead to approximately 25% dose difference at 2 cm from the source [7].

Currently there are no commercial dosimetry systems that offer a solution to this dose verification problem in brachytherapy. Some institutions have developed their own detector phantom solution, such as the Magic Phantom used for pretreatment and afterloader verification developed by Espinosa et al. [8], or pretreatment and in-vivo verification using an Electronic Portal Imaging Device [9]. However, these are highly engineered, custom-built systems not widely available in radiotherapy centers, and have not been utilized for verification measurements of dose calculation algorithms or source model verification. Furthermore, common dosimeters such as chambers, diodes, or chemical detectors such as film or gel suffer from the positional uncertainties previously mentioned.

By comparison, scanning water phantoms are ubiquitous for providing high precision relative dose measurements for external beam radiotherapy. Adapting these phantoms for brachytherapy dosimetry would serve as a convenient, low-cost solution.

The aim of this investigation was to manufacture a modular jig to enable precise and reliable dose measurements within a water phantom, around a brachytherapy source.

## 2 Method

### 2.1 Jig design and fabrication

Requirements for the jig design were established based on equipment available in the local radiotherapy department. The jig was designed to the dimensions of the PTW BeamScan water phantom (PTW, Germany) and titanium intrauterine applicator. More generally, the jig needed to: produce minimal perturbation in the scatter conditions (i.e. be mostly water equivalent with no high density components); not obstruct the mechanical movement of the water phantom arm; be reproducible and reliable.

The jig was designed using TINKERCAD (Autodesk Inc., USA) in three parts, as shown in Fig. 1: a stabilizing arm, an arm extension, and an applicator clamp. The stabilizing arm fit along one wall of the water phantom to provide two points of contact to increase the rigidity of the apparatus. It was composed of two clamps designed to match the wall thickness, connected via a 20 cm brace using dove-tail joins. Dowels were included on the underside of the clamps to fit within the screw holes in the water phantom wall to increase reproducibility of the suspended applicator.

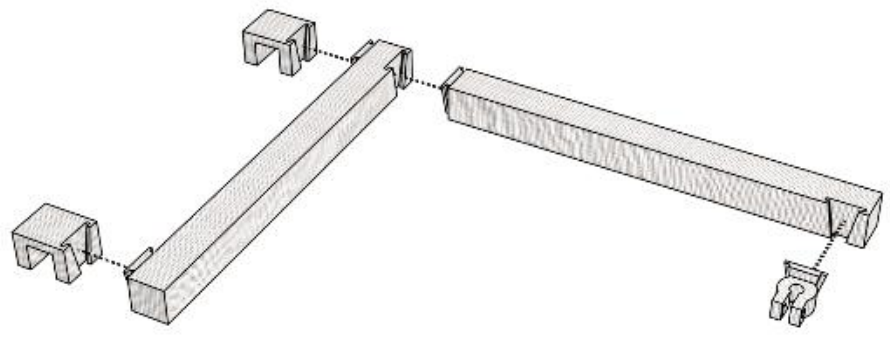

**Fig. 1.** The modular jig designed to suspend a brachytherapy applicator within a water phantom.

The arm extension used a dove-tail join to connect to the stabilizing arm, projecting another 20 cm out into the center of the tank to provide an increased scatter geometry. Another dove-tail join at this central location allowed the attachment of the applicator-specific clamp. Both the stabilizing arm and the arm extension were 3D printed in 30% infill PLA+ filament (3DFillies, Australia) to maximize the jig's rigidity and reproducibility, while minimizing manufacturing time.

The applicator clamp was designed to the diameter of a titanium intrauterine probe. This straight applicator was chosen to ensure rigidity in the experimental setup. The clamp was designed as an open "C", able to be tightened via a screw threaded through both arms of the "C" (Fig. 2). This component was 3D printed in PLAflex (3DFillies, Australia), a flexible material with a rubbery surface able to grip onto the applicator when the clamp is tightened, ensuring consistent positioning during measurements.

A Raise 3D Pro2 (Raise3D, USA) fused deposition modeling 3D printer was used to print all components of the jig, except the commercial metal screw used to tighten the applicator clamp.

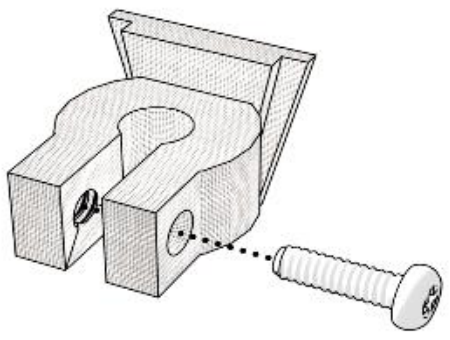

**Fig. 2.** The applicator clamp, able to be tightened using a commercial metal screw. This component is applicator specific; this example is designed for a titanium intrauterine probe.

### 2.2 Dose measurements

Relative dose distribution measurements were performed using a $^{192}$Ir Varian GammaMedPlus afterloader (Varian Medical Systems, USA). For this proof-of-concept a single dwell position was used. The applicator was suspended within the water phantom using the jig, as seen in Fig. 3. A Pinpoint 3D chamber (TN31022, PTW, Germany) was positioned underneath the applicator, where orthogonal, horizontal profiles were used to find the exact position of the source, static at a single dwell position. The chamber could then be zeroed in these two dimensions. To find the source location in the third dimension (i.e. parallel to the applicator), a vertical profile was measured 15mm from the applicator.

Once the relative position of the chamber and source position was established, two $(4 \times 4)$ cm$^2$ 2D isodose scans were measured: orthogonal to the applicator, as a radial plane; and parallel to the applicator, as an orthogonal plane. Both isodose scans were measured with an offset of 12.5 mm from the source to avoid the possibility of collision. Scans were completed at 1 mm continuous scan resolution and took between 5 and 7.5 minutes to complete.

Having established this measurement procedure and confirmed the relative positioning of the chamber and applicator, all measurements were repeated with the phantom drained of water, to provide an in-air dose distribution for comparison against a model-based dose calculation.

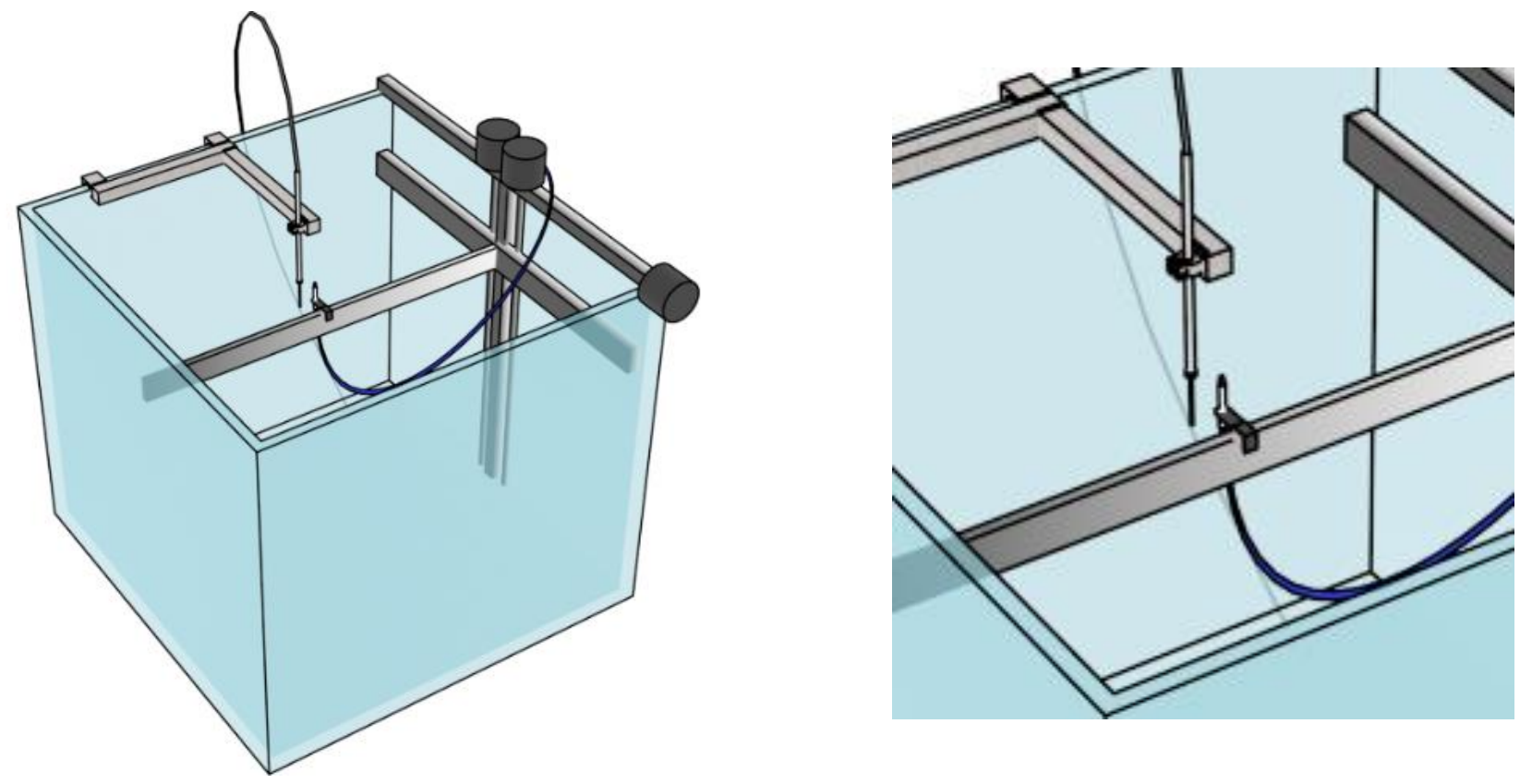

**Fig. 3.** Measurement setup. The 3D printed jig suspending the brachytherapy applicator in the water phantom, including magnified illustration of applicator and chamber arrangement.

### 2.3 Dose evaluation

Measured 2D dose planes were normalized against the maximum dose measured in each scan. These normalized dose measurements were compared against dose calculations performed using the Acuros BV algorithm in the Varian Eclipse (v13.7, Varian Medical Systems, USA) treatment planning system (TPS). These calculations were repeated in a virtual water phantom and in air, using a dose grid resolution of 1 mm.

To facilitate comparison against dose calculations, the TPS isodose planes were cropped to the same dimensions as the scan data and normalized against maximum dose values, effectively calibrating the two sets of data at a radial distance of 12.5 mm from the source. Measured and calculated isodose and dose profiles were extracted and plotted using in-house code developed using the Python programming language. Reference data published by Ballester et al. [7] were included in the comparison.

## 3 Results

The dose distribution measured around the static $^{192}$Ir source is compared with Acuros BV and Ballester et al. [7] reference data (for dose in water) in Fig. 4.

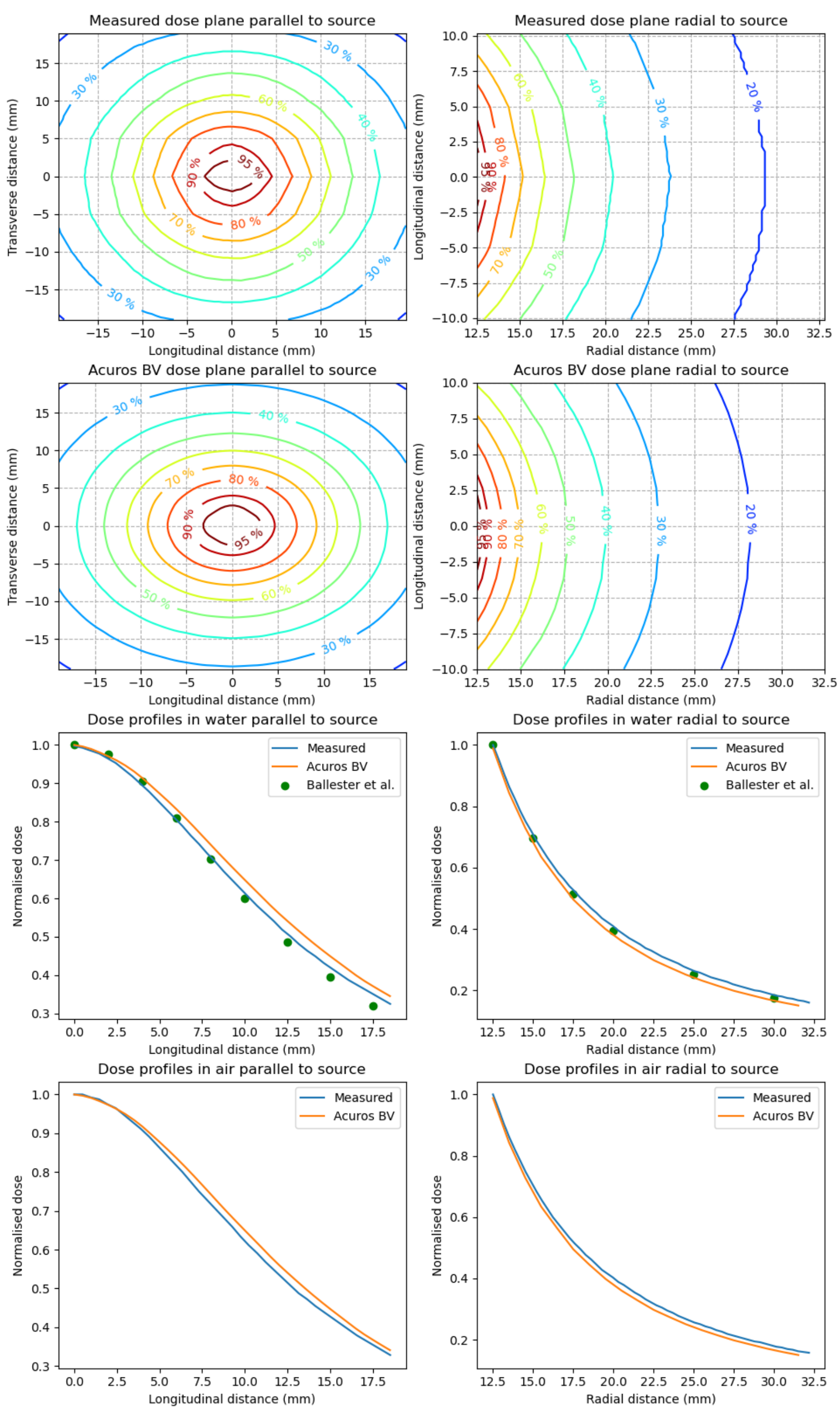

**Fig. 4.** Measured and calculated relative dose distributions

With and without water in the phantom, the normalized measured distributions agreed with those calculated within 2 mm positional uncertainty.

## 4 Discussion

The jig was designed to the dimensions of the PTW BeamScan water phantom and the titanium intrauterine probe, however, was modular enough to support alternative water phantoms and brachytherapy applicators. Moreover, the use of 3D printing and PLA material allows the jig and any adaptations to be manufactured at a low cost.

The mechanical precision of the detector positioning proved suitable for these measurements despite the high dose gradients inherent in brachytherapy. When determining the relative positions of the source and chamber, setup profiles were found to provide more accurate results when closer to the source. The importance of a high-resolution calculation grid was also noted in the comparison and analysis of results.

The comparisons between physical measurements and Acuros BV dose calculations in both water and air shown in Fig. 4 demonstrate the potential for this method to be use to verify dose calculations from model-based algorithms in media of very different densities.

This system could further be used as an alternative/verification method of source strength determination; for example, by performing precise measurement of the reference air kerma rate using the seven-distance method [10] in an unfilled phantom, or phantom-based measurements using a filled phantom.

The solution also has the potential to be used for point dose measurements for multi-dwell position, multi-channel clinical treatment plans, or to characterize dose for a single dwell position in a complex applicator containing non-water equivalent materials (e.g. shielding).

## 5 Conclusion

In this study, a 3D printed modular jig was used to convert a standard 3D scanning water phantom, conventionally used for external beam radiotherapy dosimetry, into a system for achieving precise and reliable measurements of relative dose from a brachytherapy source. Measurement results indicated the viability of this system for achieving useful comparisons with dose calculations in both water and air, indicating the potential for applications of this method in areas including verification of model-based dose calculation algorithms in different density media, performing measurements around more complex applicators and treatment geometries, and verifying estimated scatter corrections for reference dosimetry.